\documentclass[aps,11pt,preprintnumbers]{revtex4-1} 
\usepackage{amsmath}
\usepackage{amssymb}
\usepackage{latexsym}
\usepackage{ascmac}
\usepackage{graphicx}
\usepackage{color}
\usepackage{here}
\usepackage{bm}

\begin{document}

\preprint{KOBE-COSMO-19-20}

\title{Arrival Time Differences of Lensed Massive Gravitational Waves}

\author{Takuya Morita}
\email[]{183s126s@stu.kobe-u.ac.jp}
\affiliation{Department of Physics, Kobe University, Kobe 657-8501, Japan}%
\author{Jiro Soda}
\email[]{jiro@phys.sci.kobe-u.ac.jp}
\affiliation{Department of Physics, Kobe University, Kobe 657-8501, Japan}%

\begin{abstract}
It is of fundamental importance to  know the mass of gravitons.  
A simple method for constraining the graviton mass is to compare the arrival time of light and that of gravitational waves provided that both waves are simultaneously 
emitted from the same source.
To date, from observations of gravitational waves by the LIGO, the upper bound on the graviton mass $m_g$ is 
given by $m_g\lesssim 5.0 \times 10^{-23}$eV. 
However, when we compare the arrival time of light and gravitational waves,
lensing effects could be  important  for some cases. 
Moreover, in many cases, the wavelength of gravitational waves is comparable with the gravitational radius of a lens object. 
Hence, we calculate arrival time differences between electromagnetic waves and massive gravitational waves by taking into account the effect of the gravitational  wave optics. Here we take two lens models, a point mass lens and a singular isothermal sphere lens. We find that the lensing changes the arrival time difference of two waves by more than a second for the massive gravitational waves
detectable by the LISA.
\end{abstract}

\maketitle

\section{Introduction}\label{sec:intro}
In 2015, the advanced Laser Interferometer Gravitational Observatory (LIGO) detected the first event GW150914 and opened gravitational wave astronomy \cite{Abbott:2016GWdetection}. Since then, many gravitational wave events have been already observed. 
Combining observations of gravitational waves with those of
 electro-magnetic waves, we can obtain more information of the universe.
 Thus, we are now in a stage of  multi-messenger astronomy. 
In particular, it is possible to test modified gravity, namely the deviation from general relativity. For example, some of the parameters
 in scalar-tensor gravity theories are severely constrained \cite{Sakstein:2017TestingScalarTensorTheory}.

General relativity is the most successful theory where gravitons are massless.
 Recently, motivated by the accelerated expansion of the  Universe,   modified theories of gravity have been intensively investigated.
 Among them, massive gravity is an interesting possibility.  Remarkably,  a consitent theory of massive gravity theory was constructed for the first time in 2010~\cite{deRham:2010dRGTtheory}.  Irrespective of this theoretical development, it is of fundamental importance to know
the  mass of gravitons.
 In the recent paper, the LIGO collaboration announced a strong constraint on graviton mass $m_g$ as $m_g\lesssim 5.0 \times 10^{-23}$eV \cite{LIGOScientific:2019RecentResult}. 
This result has been obtained by comparison of the speed of light and that of gravitational waves.
 If  gravitational waves and electromagnetic waves are simultaneously emitted  from the same binary, the difference of arrival time shows the difference between photon and graviton masses. Assuming photon is massless, we can estimate graviton mass from the arrival time difference. There are promising sources which emit both electromagnetic waves and gravitational waves, such as supernovae, neutron star - neutron star  binary, and neutron star-Black hole mergers and actually both waves from the same binary were observed \cite{Abbott:2017NSbinary}. Thus, we expect that we can estimate the graviton mass by using this method~\cite{Larson:1999kg,Cutler:2002ef,Cooray:2003cv}. Indeed,  more stringent constraints can be  expected from the LISA in the future. 

As we mentioned, the graviton mass makes the arrival time delayed compared to
the massless case.
However, we need to consider the lensing effect for a distant source. 
It is known that lensing distorts  propagation path of massless gravitational waves in the same way as electromagnetic waves in the geometric optics limit. Moreover, gravitational waves typically have wavelength comparable to the gravitational radius of the lens objects. There, wave optics effects become important\cite{Jung:2017flg,Jung:2019fcs}. If we take into account wave optics effects~\cite{Nakamura;1999WaveOpticsGL, Nambu2012waveoptics_imageformation, Takahashi:2003WaveEffectsinGLofGWchirp}, 
gravitational waves  arrive earlier than electromagnetic waves  for massless
 gravitons~\cite{Takahashi:2016ArrivalTimeGW-EM}. 
Hence, we need to consider two competitive effects, the delay due to the graviton mass and the advance due to the lensing effect.
We must evaluate the lensing effect of a lens object 
and see if the  lensing effect is negligible or not.

Hence in this paper, we formulate the lensing of massive gravitational waves and calculate the arrival time difference in two lens models; a point mass lens and a singular isothermal sphere  lens. 
 Measuring the lens mass and the frequency of gravitational waves, we will be able to estimate the graviton mass correctly by using the LISA data in the future.

The structure of this paper is as follows. In section \ref{sec:lensed waveform}, we derive the lensed waveform in the leading order of the potential and the mass of gravitons. In section \ref{sec:Arrival Time Difference}, we derive  general formula of the arrival time difference between electromagnetic waves  and massive gravitational waves and apply the formula to two lens models. The final section \ref{sec:discuss} is devoted to the conclusion.
In this paper,   we use units $G=c=\hbar=1$. 

\section{Lensing of massive gravitational waves}\label{sec:lensed waveform}
In this section,  we derive the general formula for a lensed waveform of massive gravitational waves. 

If we ignore polarization,  the equation of motion of gravitational waves coincides with  that of a scalar field.
It is legitimate to assume that massive gravitational waves can be described by the equation of 
a  massive  scalar field $\phi(t, \vec{r})$ 
\begin{align}
	\left[\frac{1}{\sqrt{-g}}\partial_{\mu}\left(\sqrt{-g}g^{\mu\nu}\partial_{\nu}\right)-m_{g}^{2}\right]
		\phi(t, \vec{r})=0\label{eq:2.00}\ ,
\end{align}
where $m_g$ is the mass of gravitons and  $\vec{r}$ is a position vector. 

In Newtonian approximation, the metric around a lens object reads
\begin{align}
	ds^{2}&= -\left(1+2U(\vec{r})\right)dt^{2}+\left(1-2U(\vec{r})\right) d\vec{r}^{\ 2}\label{eq:2.0}   \ ,
\end{align}
where $U(\vec{r})$ is the gravitational potential. From the Poisson equation
\begin{align}
	\nabla^2 U(\vec{r})=4\pi G\rho(\vec{r})\ ,
\end{align}
we can obtain the gravitational potential
\begin{align}
	U(\vec{r})=-\int\frac{d^3\vec{r}'}{|\vec{r}-\vec{r}'|}\rho(\vec{r}')  \  ,
\end{align}
where  $\rho(\vec{r})$ is the mass density.
Note that we can assume $U(\vec{r})\ll1$.
Using  the metric  \eqref{eq:2.0} in eq. \eqref{eq:2.00}, we obtain
\begin{align}
	\left[\omega^{2}+\nabla^{2}-m_{g}^{2}\right]\tilde{\phi}(\omega,\vec{r})=4\omega^{2}U(\vec{r})
		\left(1-\frac{m^2_g}{2\omega^2}\right)\tilde{\phi}(\omega,\vec{r})\label{eq:2.1}\ ,
\end{align}
where $\tilde{\phi}(\omega, \vec{r})$ is  the  Fourier transformation of $\phi(t, \vec{r})$ with respect to the time. $\omega$ is an angular frequency and $\nabla^2\equiv\partial_i\partial^i$.

\begin{figure}
\begin{centering}
\includegraphics[scale=1.1]{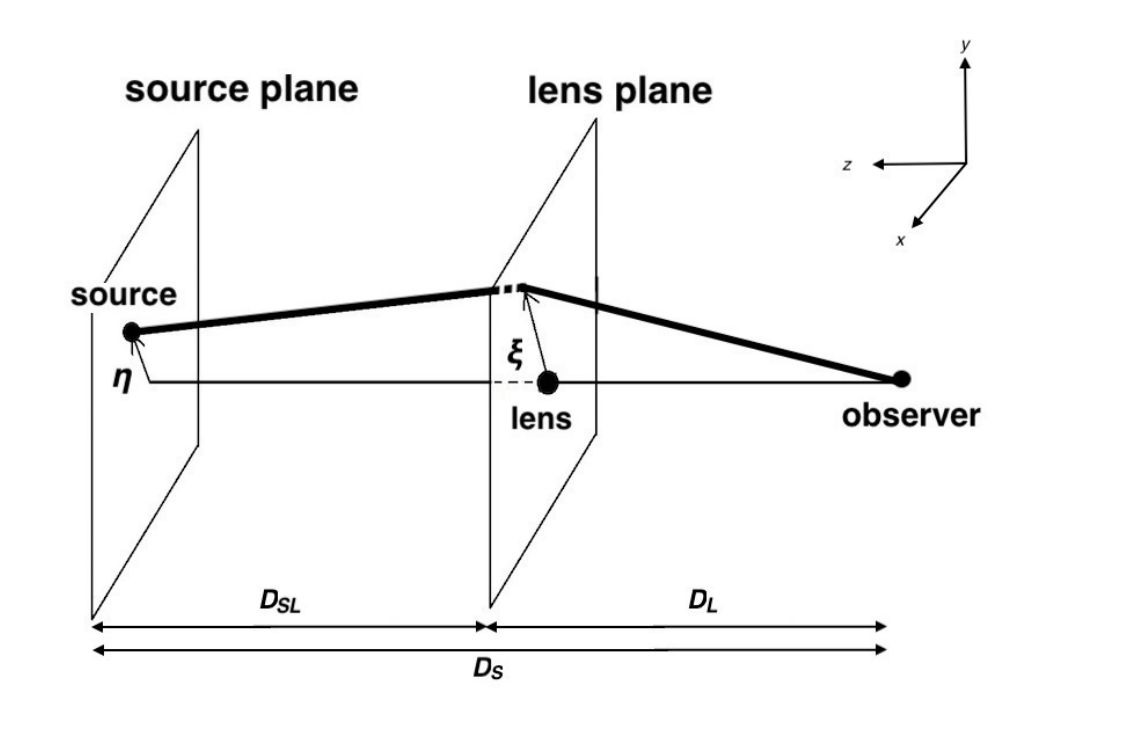}
\par\end{centering}
\caption{
Schematic picture of gravitational wave lensing in the geometric optics limit. Observer, lens, and source positions are $(\bm{0}, 0), (\bm{0}, D_L), (\bm{\eta}, D_S)$, respectively. We assume the thin lens approximation is valid. Therefore, gravitational waves from the source are reflected only at ($\bm \xi$, $D_L$). 
\label{fig:wave reflection plane}}
\end{figure}

The right hand side of eq.\eqref{eq:2.1} represents lensing effect which changes the  path of wave propagation.
Since the range of potential is small compared to the size of the system, we use the thin lens approximation, namely, 
we assume that  the lens potential $U(\vec{r})$ distorts the  path of gravitational waves only once near the lens.
In Fig.\ref{fig:wave reflection plane}, we depicted a schematic picture of
gravitational wave lensing in the geometric optics limit. We take $z$ axis as the direction from the observer to the lens object. From now on, we represent three dimensional coordinates $(x, y, z)$ 
as $(\bm{X}, z)$, where $\bm{X}$ is a two dimensional position vector in the $x-y$ plane, namely, $\bm{X}\equiv(x, y)$. The observer defines the origin $(\bm{0}, 0)$, the lens obeject and the source are located at $(\bm{0}, D_L)$ and $(\bm{\eta}, D_S)$, respectively.  We call $z=D_L$ lens plane and $z=D_S$ source plane. 
Here,  $\bm{\eta}$ is a two dimensional position vector of the source 
on the source plane. 
Under the thin lens approximation, gravitational waves are bended only once at the lens plane $z=D_L$. Namely, gravitational waves are emitted from the source ($\bm\eta$, $D_S$),  refracted at the position ($\bm\xi$, $D_L$), and finally arrive at the observer ($\bm0$, 0). Note that a 2 dimensional position vector on the lens plane $\bm{\xi}$ is determined by the lens equation. We also represent the distance between the lens plane and the source plane by $D_{SL}$.

The solution of eq.\eqref{eq:2.1} can be written as $\tilde{\phi}(\omega,\vec{r})=\tilde{\phi}_0(\omega,\vec{r})+\tilde{\phi}_1(\omega,\vec{r})$, where $\tilde{\phi}_0(\omega,\vec{r})$ is the unlensed solution obtained by setting  $U(\vec{r})=0$ and $\tilde{\phi}_1(\omega,\vec{r})$ is the lensed solution with the thin lens approximation. The unlensed solution is given by~\cite{Takahashi:2005ScatteringGWbyGL} 
\begin{align}
	\tilde{\phi}_{0}(\omega,\vec{r})= & \frac{\exp\left[i\sqrt{\omega^2-m_{g}^{2}}\int^{D_{S}}dr\right]}{4\pi |\vec{r}-\vec{r}_s|} \ .
\end{align}
Using the unlensed solution, we can deduce the lensed solution as
\begin{align}
	\tilde{\phi}_{1}(\omega, \vec{r})= & -\frac{\omega}{2\pi i}\int d\vec{r}'
		\frac{\exp\left[i\sqrt{\omega^2-m_{g}^{2}}\int^{D_{L}}dr\right]}
		{\left|\vec{r}-\vec{r}'\right|} \nonumber\\
	&\times2i\omega U(\vec{r}')\left(1-\frac{m_{g}^{2}}{2\omega^{2}}\right) 
		\frac{\exp\left[i\sqrt{\omega^2-m_{g}^{2}}\int_{D_{L}}^{D_{S}} dr\right]}
		{4\pi|\vec{r}'-\vec{r}_s|}\ ,
\end{align}
 where $\int^{D_i} dr$ ($i$=L, SL, S) are the integrals along the propagation path and $\vec{r}_s$ is the 3 dimensional source position.

Now, we introduce the lensing amplification factor $F$ as 
\begin{align}
	F\equiv\frac{\tilde{\phi}_0+\tilde{\phi}_1}{\tilde{\phi}_0}\ .
\end{align}
Using the thin lens approximation and  $\xi, \eta\ll D_i$, we obtain
\begin{align}
	F\simeq & \frac{\omega}{2\pi i}\frac{D_{S}}{D_{L}D_{SL}}\int d^{2}\bm{\xi}
		e^{-\frac{m_{g}^{2}}{2\omega^{2}}} \nonumber\\
		 & \times\exp\left[i\omega\left(1-\frac{m_{g}^{2}}{2\omega^{2}}\right)\left\{
		\frac{D_{S}}{D_{L}D_{SL}}\frac{\bigg|\bm{\xi}-\frac{D_{L}}{D_{S}}\bm{\eta}\bigg|^{2}}{2}
		-\psi(\bm{\xi}) \right\}\right]\label{eq:2.3}\ ,
\end{align}
where we assumed $m_g\ll\omega$ and took up to the second order of $m_g/\omega$. 
Here, we defined a two dimensional gravitational potential $\psi(\bm\xi)$
\begin{align}
	\psi(\bm\xi)\equiv \int dz \ 2U(\bm\xi, z)\ .
\end{align}
Note that  $\psi(\bm\xi)$ has a  constant of integration. 
For the unlensed case, i.e., $U(\vec{r})=0$, we should have $F=1$.
Hence, we can determine the constant of integration so that $F=1$ when $U(\vec{r})=0$.
This formula reduces to the known result in the limit $m_{g}=0$ \cite{Takahashi:2016ArrivalTimeGW-EM}. 
The first and second terms in the exponent are geometrical time delay and 
 the Shapiro time delay, respectively \cite{Shapiro:1964ShapiroDelay}. 

It is convenient to define the phase function of the amplification factor as
\[
	T(\omega)\equiv\frac{1}{i\omega\left(1-\frac{m^2_g}{2\omega^2}\right)}\ln\left(\frac{F}{|F|}\right)\ .
\]
By using $\tilde{\phi}=F\tilde{\phi}_0$ and $\phi=\int d\omega \exp[-i\omega t]\tilde{\phi}$, the gravitational wave form $\phi$ can be written as
\begin{align}
	\phi\propto\int d\omega e^{i\omega\left(1-\frac{m^2_g}{2\omega^2}\right)(D_S+T(\omega))-i\omega t}\label{eq:2.4}\ .
\end{align}
In the next section, we derive the propagation time of gravitational waves with the phase function and apply the formula to the two lens models, Point Mass lens (PM) and Singular Isothermal Sphere lens (SIS).

\section{Arrival Time Differences}\label{sec:Arrival Time Difference}
In this section, we compute the arrival time difference for the point mass lens and the singular isothermal sphere lens. The behavior of lensing amplification factor depends on the frequency $\omega$ of gravitational waves. Hence, we carefully treat the
 amplification both for the  geometrical optics limit and the wave optics limit, separately.

\subsection{Group Velocity}\label{subsec:GroupVelocity}
In this subsection, we derive the  group velocity of  massive gravitational waves.
Consider gravitational waves with a gaussian distribution  of frequencies  centered at $\omega_0$ and the dispersion $\sigma^2$.
The  exponent of eq. \eqref{eq:2.4} can be expanded  around $\omega_0$ as
\begin{align}
	&i\left(\omega-\frac{m^2_g}{2\omega}\right)(D_S+T(\omega))-i\omega t\nonumber\\
	\simeq&i\left(\omega_0-\frac{m^2_g}{2\omega_0}\right)(D_S+T(\omega_0))-i\omega_0 t\nonumber\\
	&+i\left(\left(1+\frac{m^2_g}{2\omega^2_0}\right)(D_S+T(\omega_0))
		+\left(\omega_0-\frac{m^2_g}{2\omega_0}\right)T'(\omega_0)-t\right)\delta\omega\nonumber\\
	&+i\left\{-\frac{m^2_g}{\omega^3_0}(D_S+T(\omega_0))
		+2\left(1+\frac{m^2_g}{2\omega^2_0}\right)T'(\omega_0)
		+\left(\omega_0-\frac{m^2_g}{2\omega_0}\right)T''(\omega_0)\right\}\delta\omega^2\ ,
\end{align}
where a dash represents a partial derivative with respect to the frequency. 
Denoting  the coefficients of $\delta\omega^2$ as $N$,  we have
\begin{align}
	\phi\propto&\int d(\delta\omega)\frac{1}{\sqrt{2\pi\sigma^2}}e^{-\frac{\delta\omega^2}{2\sigma^2}}
		\times e^{i\omega_0\left(1-\frac{m^2_g}{2\omega^2_0}\right)(D_S+T(\omega_0))-i\omega_0 t}\nonumber\\
	&\times e^{i\left(\left(1+\frac{m^2_g}{2\omega^2_0}\right)(D_S+T(\omega_0))
		+\left(\omega_0-\frac{m^2_g}{2\omega_0}\right)T'(\omega_0)-t\right)\delta\omega
		+iN\delta\omega^2}  \ .   \nonumber 
\end{align}
After integration, we can deduce the group velocity of gravitational  waves  as
\begin{align}
	\propto&e^{i\omega_0\left(1-\frac{m^2_g}{2\omega^2_0}\right)(D_S+T(\omega_0))
		-i\omega_0 t}\nonumber\\
	&\times e^{\frac{1}{4\left(iN-\frac{1}{2\sigma^2}\right)}\left(\left(1+\frac{m^2_g}{2\omega^2_0}\right)(D_S+T(\omega_0))
		+\omega_0\left(1-\frac{m^2_g}{2\omega^2_0}\right)T'(\omega_0)-t\right)^2}\ .
\end{align}
The last factor describes the time evolution of  wave packets. 
Thus,  the propagation time is given by the formula
\begin{align}
	t=\left(1+\frac{m^2_g}{2\omega^2_0}\right)(D_S+T(\omega_0))
		+\omega_0\left(1-\frac{m^2_g}{2\omega^2_0}\right)T'(\omega_0)\label{eq:g.6}\ .
\end{align}

In the  geometric optics limit, we can use Fermat's Principle, 
namely, waves reflect at the stationary point of the phase $\bm{\xi}=\bm{\xi}(\bm{\eta})$
\begin{align}
	\partial_{\bm{\xi}} \left[i\omega\left(1-\frac{m^2_g}{2\omega^2}\right)
		\left(\frac{D_{S}}{D_{L}D_{SL}}
		 \frac{\bigg|\bm{\xi}-\frac{D_{L}}{D_{S}}\bm{\eta}\bigg|^{2}}{2}-\psi(\bm{\xi})\right)\right]=0\ .
\end{align}
Thus we can calculate the phase function in the geometric optics limit $T_{\rm geo}$
\begin{align}
	T_{\rm geo}& =\frac{D_{S}}{D_{L}D_{SL}}\frac{\bigg|\bm{\xi}(\bm{\eta})
		-\frac{D_{L}}{D_{S}}\bm{\eta}\bigg|^{2}}{2}-\psi(\bm\xi(\bm\eta))
	 \label{eq:3.2}\ .
\end{align}
This result is independent of $\omega$, so reproduces eq.(20) - (23) in the previous work (in the subhorizon limit $a(t)=1$) \cite{Baker:2016MultimessengerTimeDelay}. 
In the wave optics limit, on the other hand, we must evaluate the integral in 
eq.\eqref{eq:2.3} accurately. 

In a realistic situation,  the frequency of gravitational waves is in the wave optics range and the frequency of electromagnetic waves  is in the geometrical optics range. 
Hence, there appears the arrival time difference. 
Now, we can define the arrival time difference between electromagnetic waves
 and gravitational waves  as
\begin{align}
	\Delta t&\equiv (\text{propagation time of EMWs})
	-(\text {propagation time of GWs})\label{eq:3.3} \ .
\end{align}
Here, we assumed the mass of a photon exactly vanishes.
In the following subsections, we calculate $\Delta t$ for the point mass lens and 
the singular isothermal sphere lens. As an illustration, we consider a single lens object at the center of our galaxy and gravitational waves from the opposite side of the center of our galaxy. In this paper, we assume electromagnetic waves and gravitational waves are emitted from the source at the same time, the lens object is around the center of our galaxy ($D_L=8$ kpc) and $D_S\simeq 1$Gpc.

\subsection{Point Mass Lens}\label{subsec:PM}
As the first example, we consider a point mass  lens. Since the lens is located  at $(\bm{0}, D_L)$, 
the mass density is given by
\begin{align}
	\rho(\vec{r})&=M_L\delta^3(\vec{r}-(\bm{0},D_L))\label{eq:4.1}\ ,
\end{align}
where $M_L$ is the mass of a lens object. 
It is easy to get the lens potential
\begin{align}
	U(\vec{r})&=-\frac{M_L}{|\vec{r}-(\bm{0},D_L)|}\ .
\end{align}
After integration, we obtain the  two dimensional potential
\begin{align}
	\psi(\bm\xi)&=4M_L\ln|\bm\xi|     \label{eq:4.2}\ ,
\end{align}
Inserting eq. \eqref{eq:4.2} into eq. \eqref{eq:2.3}, we can deduce the  amplification factor.

Using the Einstein radius $\xi_0$ in this model defined by  (for example, see \cite{Dodelson:2016GravitationalLensing})
\begin{align}
	\xi_0=\sqrt{4M_L\frac{D_LD_{SL}}{D_S}} \ ,  \label{eq:4.5}
\end{align}
we can define dimensionless positions $\bm{y}$ and $\bm{u}$ as
\begin{align}
	\begin{cases}
	\bm{y}\equiv&\frac{\bm\xi}{\xi_0}    \ , \\
	\bm{u}\equiv&\frac{D_L}{D_S}\frac{\bm\eta}{\xi_0}   \ .\label{eq:4.4}
	\end{cases}\ 
\end{align}
Using these variables, we can rewrite the amplification factor $F$ as
\begin{align}
	F=  \frac{We^{-\frac{m_{g}^{2}}{2\omega^{2}}}}{2\pi i}\int d^{2}\bm{y} 
	  \exp\left[iW\left(1-\frac{m_{g}^{2}}{2\omega^{2}}\right)\left( \frac{|\bm{y}-\bm{u}|^{2}}{2}-\ln|\bm{y}|\right)\right]    \label{eq:4.6}\ ,
\end{align}
where $W\equiv4M_L\omega$ is the dimensionless frequency. 

In the geometric optics limit ($1\ll W$), we can apply the Fermat's Principle to the integral  \eqref{eq:4.6}. 
Using variables \eqref{eq:4.4}, the Fermat's Principle is written as $\partial_{\bm{y}}\left( \frac{|\bm{y}-\bm{u}|^{2}}{2}-\ln|\bm{y}|\right)=0$. 
As the simplest setup, we assume that $\bm y$ is parallel to $\bm u$.
Thus, the magnitude $y(u)\equiv|\bm{y}(\bm{u})|$ reads
\begin{align*}
	y(u)=\frac{u\pm \sqrt{u^2+4}}{2}\ .
\end{align*}
Here, the plus and minus signs correspond to outer and inner images of lensing.  
Inserting this into eq. \eqref{eq:3.2}, the phase function becomes
\begin{align*}
	T_{\text{PM, geo},\pm} =4M_L\left(\frac{u^2+2\mp u\sqrt{u^2+4}}{4}
		-\ln\bigg|\frac{u\pm\sqrt{u^2+4}}{2}\bigg|\right)\ .
\end{align*}
Thus, the propagation time in the geometric optics limit is
\begin{align}
	t_{\text{PM, geo},\pm} =\left(1+\frac{m_{g}^{2}}{2\omega^{2}}\right)
		\left\{D_{S}+4M_L\left(\frac{u^2+2\mp u\sqrt{u^2+4}}{4}
		-\ln\bigg|\frac{u\pm\sqrt{u^2+4}}{2}\bigg|\right)\right\}\label{eq:4.8}\ .
\end{align}
Hereafter, we set $u=0.8$ for simplicity.

In the wave optics limit ($W\ll1$), on the other hand, the amplification factor becomes
\begin{align}
	F\propto\exp\left[\frac{iW\left(1-\frac{m^2_g}{2\omega^2}\right)}{2}
		\left\{\ln\left(\frac{W\left(1-\frac{m^2_g}{2\omega^2}\right)}{2}\right)+\gamma\right\}\right]\ ,
\end{align}
where $\gamma$ is the Euler constant. So the phase function can be evaluated as 
\begin{align*}
	T_{\text{PM, wave}}(\omega)
		=2M_L\left\{\ln\left(2M_L\omega\left(1-\frac{m^2_g}{2\omega^2}\right)\right)+\gamma\right\}\ ,
\end{align*}
 and therefore the propagation time is given by
\begin{align}
	 t_{\text{PM, wave}}=\left(1+\frac{m^2_g}{2\omega^2_0}\right)\left\{D_S
	 	+2M_L\left[\ln\left(2M_L\omega\left(1-\frac{m^2_g}{2\omega^2}\right)\right)
	 	+\gamma+1\right]\right\}\label{eq:4.61}\ .
\end{align}

Before computing $\Delta t$ in the LISA band, we check if the wave optics effect is relevant  or not in the LIGO band. 
From the formula
\begin{align*}
	W= 2 \left( \frac{M}{10^4 M_{\odot}}  \right) \left( \frac{\omega}{10 {\rm Hz}}  \right)    \ ,
\end{align*}
we see, for the   LIGO frequency band $10-10^3$Hz, lens objects with $10^4 M_\odot<M_L$ lead to $1< W$. Namely, the wave optics effect is not relevant.
For the lens mass $M_L  < 10^4 M_\odot$, the time advance due to the wave effect 
\begin{align*}
	\left|2M_L\right|\sim 0.1\left( \frac{M_L}{10^4 M_{\odot}}  \right) {\rm s}    
\end{align*}
is small.
We can also evaluate the time delay due to the graviton mass as 
\begin{align*}
	\left( \frac{m_g}{\omega}\right)^2 D_S = 10^{-7}  \left( \frac{m_g}{10^{-26} {\rm eV}}  \right)
		 \left( \frac{10 {\rm Hz}}{\omega}  \right)^2 \left( \frac{D_S}{1{\rm Gpc}}  \right)   {\rm s}    \ .
\end{align*}
This  delay is also negligible for the LIGO band.
Thus,  both the wave optics effect and the effect of graviton mass  for the LIGO band is negligible.
For the LISA band, on the other hand, the wave effect is relevant for a lens mass smaller than $10^8 M_\odot$. Moreover, both the time advance due to the wave effect and the time delay due to the graviton mass becomes sizable.
For example, if the graviton mass is $10^{-26}$ eV and the lens mass is $10^4 M_\odot$, the arrival time difference is on the order of a second.
Of course, the detail depends on the lens model, which we are now investigating.

\begin{figure}
\begin{centering}
\includegraphics[scale=0.55]{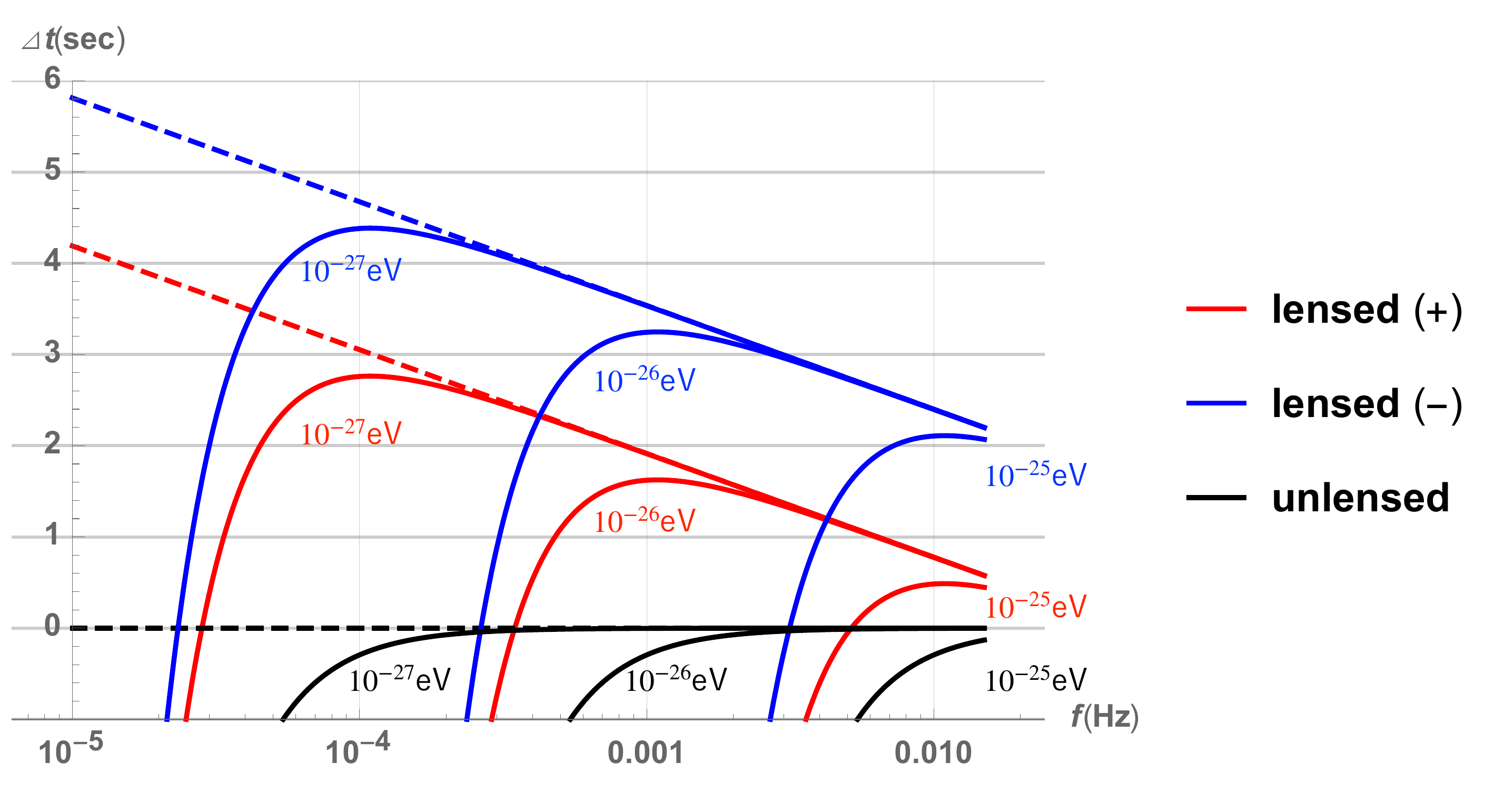}
\par\end{centering}
\caption{
Arrival time differences for a point mass lens in the LISA band. For drawing red and blue lines, we assumed  $M_L=5\times10^4M_\odot$.
We also  depicted black lines corresponding to unlensed cases $M_L=0$. Plus and minus signs correspond to red and blue. Horizontal axes represents a frequency $f$ 
ranging $W=6\times10^{-5} - 6\times10^{-2}\ll1$ (wave optics limit). Each solid line represents the graviton mass in the range $m_{g}=10^{-25} - 10^{-27}$eV 
and the dashed line represents $m_g=0$. From this figure, we can see that the arrival time difference due to the lensing effect is on the order of $ \mathcal{O}(1)$, so we must consider 
the lensing effect in the analysis of the LISA data.
\label{fig:Arrival-time-difference}}
\end{figure}

Now, we see $\Delta t$ in the  LISA band. Figure \ref{fig:Arrival-time-difference} represents $\Delta t$ for three cases. Red and blue lines represent lensed cases with $M_L =5\times10^4M_\odot$, and black lines represent unlensed cases. The frequency band in this figure corresponds to $W=6\times10^{-5} - 6\times 10^{-2}$. 
Each solid line represents the graviton mass in the range $m_{g}=10^{-25} \sim 10^{-27}$eV and the dashed line represents $m_g=0$.

Here the gaps between red (blue) and black lines in this figure is on the  order of a second due to the lensing effect. 
The error in the LISA band is of the same order~ \cite{Will:1998GravitonMassbyGW}, so it can be  important  for the LISA data. 
Therefore, we must consider the lensing effect when we estimate the  graviton mass using the LISA data. In principle, by measuring the arrival time difference $\Delta t$ and the frequency $\omega$, we can estimate the graviton mass with this figure.

Here we mention about the probability of lensing. If we take the density as critical density of the universe $\rho=\rho_c=10^{-29} \text{g/cm}^3=130M_\odot/\text{kpc}^3$ and source position as $D_S=1$Gpc, optical depth, which is the number of lens objects between observer and the source, become
\begin{align}
	\tau&=\int^{D_S}_0 dD_L\ \pi\xi^2_0\frac{\rho(D_L)}{M_L}\nonumber\\
	&=1.7\times10^3M_\odot \text{kpc}^{-3}\int^{D_S}_0dD_L\frac{D_L(D_S-D_L)}{D_S}\simeq1.4\times10^{-2}\ .
\end{align}
Therefore, gravitational waves is probable to be lensed.

\subsection{Singular Isothermal Sphere Lens}\label{subsec:SIS}
For  a singular isothermal sphere lens model,  we take
 the mass density
\begin{align}
	\rho(\vec{r})&=\frac{\sigma^{2}}{2\pi |\vec{r}^{2}|}\label{eq:4.10}\ ,
\end{align}
where $\sigma$ is a velocity dispersion.
We can deduce the gravitational potential
\begin{align}
	U(\vec{r})&=2\sigma^{2}\ln\left(\frac{|\vec{r}|}{r_{0}}\right)\label{eq:4.11}\ ,
\end{align}
where $r_0$ is the cut off scale.  
It is easy to get the  two dimensional gravitational potential 
\begin{equation}
	\psi(\bm\xi)=4\pi\sigma^{2}|\bm\xi|\ .
\end{equation}
The Einstein radius and the lens mass in this model are given by
\begin{equation}
	\xi_{0}=\frac{4\pi\sigma^{2}D_{L}D_{SL}}{D_{S}}\label{eq:4.17}\ ,
\end{equation}
and 
\begin{equation}
	M_L=\frac{4\pi^2\sigma^4D_LD_{SL}}{D_S}   \ , \label{eq:4.18}
\end{equation}
respectively.

\begin{figure}
\begin{centering}
\includegraphics[scale=0.55]{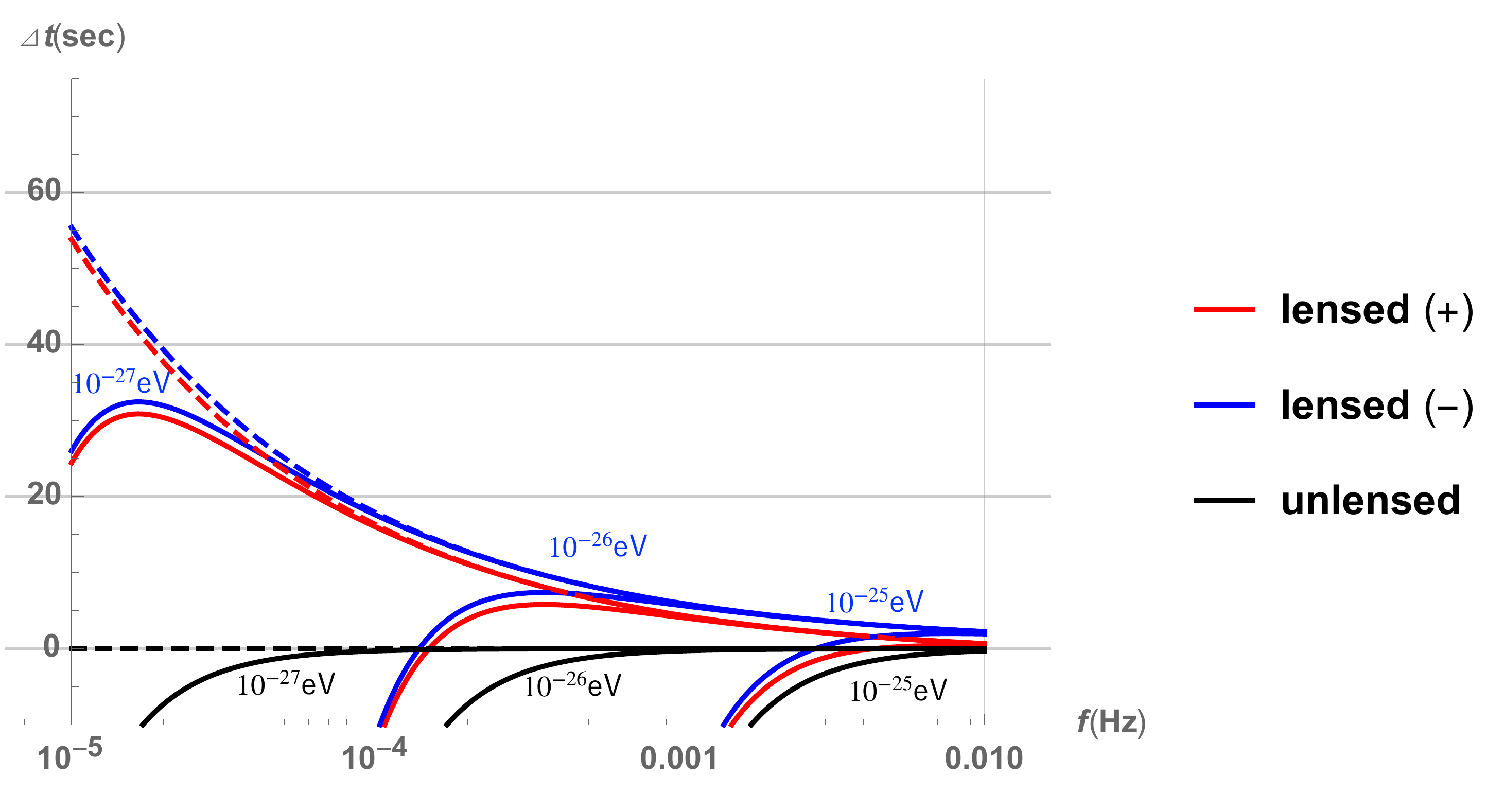}
\par\end{centering}
\caption{
Arrival time difference of SIS lens in the LISA frequency band. Axes are same as fig \ref{fig:Arrival-time-difference}. Here we set $M_L=5\times10^4M_\odot$ and $\sigma\simeq 80$km/sec, so plot range of $W$ is same as fig \ref{fig:Arrival-time-difference}. Each solid line differs in $m_{g}=10^{-24}\sim10^{-27}$eV and dashed lines represent $m_g=0$. 
\label{fig:Arrival-time-difference_SIS}}
\end{figure}

The amplification factor $F$ in this model  reads \cite{Matsunaga:2006SISderivation}
\begin{align}
	F\simeq \frac{We^{-\frac{m_{g}^{2}}{2\omega^{2}}}}{2\pi i}\int d^{2}\bm{y} 
		 \exp\left[iW\left(1-\frac{m_{g}^{2}}{2\omega^{2}}\right)
		 \left\{\frac{\bigg|\bm{y}-\bm{u}\bigg|^{2}}{2}-|\bm y| \right\}\right]\label{eq:4.14}\ ,
\end{align}
where we  used $W=4M_L\omega$.

In the geometric optics limit ($1\ll W$), we can apply Fermat's Principle and get 
\begin{align*}
	y(u)=u\pm1\ .
\end{align*}
Hence, the phase function is obtained as
\begin{align*}
	T_{\text{SIS, geo}, \pm}=4M_L\left(\mp u-\frac{1}{2}\right)
\end{align*}
and the propagation time is given by
\begin{align}
	t_{\text{SIS, geo}, \pm}=&\left(1+\frac{m_{g}^{2}}{2\omega^{2}}\right)\left\{D_S+4M_L\left(\mp u-\frac{1}{2}\right)\right\}\ .
		\label{eq:4.16}
\end{align}

In wave optics limit ($W\ll1$), the phase function is given by
\begin{align*}
	T_{\text{SIS, wave}}(\omega)=4M_L\left\{-\frac{\sqrt{\pi}}{2}
		\left(4M_L\omega\left(1-\frac{m^2_g}{2\omega^2}\right)\right)^{-1/2}
		-\left(1-\frac{\pi}{4}\right)\right\}
\end{align*}
and the propagation time can be evaluated as
\begin{align}
	t_{\text{SIS, wave}}\simeq&\left(1+\frac{m^2_g}{2\omega^2}\right)\left[D_S
		+4M_L\left\{-\frac{\sqrt{\pi}}{4}\left(4M_L\omega
		\left(1-\frac{m^2_g}{2\omega^2}\right)\right)^{-1/2}
		-\left(1-\frac{\pi}{4}\right)\right\}\right]\label{eq:4.15}\ .
\end{align}
In Fig.\ref{fig:Arrival-time-difference_SIS}, we  depicted the arrival time difference $\Delta t$ using the formulas  \eqref{eq:4.16} and \eqref{eq:4.15}. 
The red and blue lines are lensed cases with $\sigma\simeq80$ km/sec. The time difference $\Delta t$ in the singular isothermal sphere lens is larger than that in the point mass lens. 

Thus, we have shown that the arrival time difference of gravitational waves and electro-magnetic waves depends on lens models.

\section{Conclusion}\label{sec:discuss}

Since it is important to  know the mass of gravitons, we have studied   
a simple method for constraining the graviton mass. The method is to compare the speed of light 
and that of gravitational waves by
assuming the simultaneous emission of both waves from the same source.
What we need is to compare the arrival time difference. 
However, when we compare the arrival time of light and gravitational waves,
lensing effects could be  important. 
Moreover, wave optics would be relevant for lensing of gravitational waves. 
In fact, in many cases, the wavelength of gravitational waves is comparable with the gravitational radius of a lens object. 
Thus, we have derived the arrival time difference between lensed electromagnetic waves and massive gravitational waves
using the wave optics in gravitational lensing. 
We used a point mass lens and a singular isothermal sphere lens. 
We have shown that when waves in the LISA band pass near a massive object $M_L\sim 5\times10^4M_\odot$, the arrival time difference is about $\mathcal{O}(1)$ sec for a point mass lens and about $\mathcal{O}(10)$ sec for a singular isothermal sphere lens. Moreover, we showed that it is likely enough to observe the lensed gravitational waves. 
Therefore, when we estimate graviton mass from the LISA data in future, the lensing effect must be considered. 

\begin{acknowledgments}
J.~S. was in part supported by JSPS KAKENHI
Grant Numbers JP17H02894, JP17K18778, JP15H05895, JP17H06359, JP18H04589.
J.~S. was also supported by JSPS Bilateral Joint Research
Projects (JSPS-NRF collaboration) `` String Axion Cosmology.''
\end{acknowledgments}

\bibliographystyle{apsrev4-1}
\bibliography{BibTexMast.bib}

\end{document}